\documentclass{IEEEcsmag}

\usepackage[colorlinks,urlcolor=blue,linkcolor=blue,citecolor=blue]{hyperref}
\expandafter\def\expandafter\UrlBreaks\expandafter{\UrlBreaks\do\/\do\*\do\-\do\~\do\'\do\"\do\-}
\usepackage{upmath,color}

\jvol{XX}
\jnum{XX}
\paper{8}
\jmonth{December}
\jname{Publication Name}
\jtitle{Synergizing Monetization, Orchestration, and
Semantics in Computing Continuum}
\pubyear{2025}

\begin{document}

\sptitle{Article Type: Regular}

\title{Synergizing Monetization, Orchestration, and Semantics in the Computing Continuum}

\author{Chinmaya Kumar Dehury}
\affil{IISER Berhampur, Odisha, India}

\author{{}Lauri Lovén} 
\affil{Center for Ubiquitous Computing, University of Oulu, Oulu, Finland}

\author{Praveen Kumar Donta}
\affil{Computer and Systems Sciences, Stockholm University, Stockholm, Sweden}

\author{Ilir Murturi}
\affil{Department of Mechatronics, University of Prishtina, Prishtina, Kosova}

\author{Schahram Dustdar}
\affil{Distributed Systems Group, TU Wien, Vienna, Austria. and ICREA, Barcelona, Spain}

\markboth{THEME/FEATURE/DEPARTMENT}{THEME/FEATURE/DEPARTMENT}

\begin{abstract}\looseness-1
Industry demands are growing for hyper-distributed applications that span from the cloud to the edge in domains such as smart manufacturing, transportation, and agriculture. Yet today's solutions struggle to meet these demands due to inherent limitations in scalability, interoperability, and trust. In this article, we introduce HERMES (Heterogeneous Computing Continuum with Resource Monetization, Orchestration, and Semantic) – a novel framework designed to transform connectivity and data utilization across the computing continuum. HERMES establishes an open, seamless, and secure environment where resources, from cloud servers to tiny edge devices, can be orchestrated intelligently, data and services can be monetized in a distributed marketplace, and knowledge is shared through semantic interoperability. By bridging these key facets, HERMES lays a foundation for a new generation of distributed applications that are more efficient, trustworthy, and autonomous.
\end{abstract}

\maketitle

\chapteri{E}dge computing was envisioned over a decade ago as a means to decentralize computation away from cloud data centers toward the network edge, closer to where data is generated~\cite{shi2016edge}. The goal was to reduce latency, save bandwidth, and improve data privacy by processing information locally. Major industry players even proposed \emph{fog computing} architectures to extend cloud capabilities outwards, installing mini cloud-like resources at/near the edge of the network~\cite{bonomi2012fog}. In practice, however, most IoT devices and edge systems today remain significantly underutilized, often acting merely as sensors that relay data to the cloud, rather than performing on-site analytics or decision-making \cite{murturi2021decentralized}. %

Three major challenges have hindered the fulfillment of the edge computing promise. First, many current solutions lead to \textit{vendor lock-in} and poor interoperability \cite{pujol2023edge}. Cloud providers have addressed edge computing as an extension of their own proprietary cloud infrastructure~\cite{vano2023cloud}, creating siloed deployments where edge devices must run vendor-specific stacks and cannot easily mix with third-party services. Second, there are significant \textit{orchestration gaps} – cloud-native orchestration platforms (e.g., Kubernetes) struggle in resource-constrained, diverse edge environments and often assume reliable connectivity and homogeneous infrastructure. Managing applications across thousands of far-flung devices, each with distinct capabilities, remains a significant challenge. Third, mechanisms for \textit{trust, data sharing, and monetization} at the edge are still not well researched. Companies are hesitant to expose or trade IoT data due to security and privacy concerns, and there are few widely adopted frameworks to incentivize the sharing of computational resources or data streams between organizations~\cite{christidis2022decentralized}. Although research has explored blockchain-based IoT data marketplaces, real-world deployments of such marketplaces are still nascent.

\begin{figure*}[h]
\centering
\includegraphics[width=0.9\textwidth]{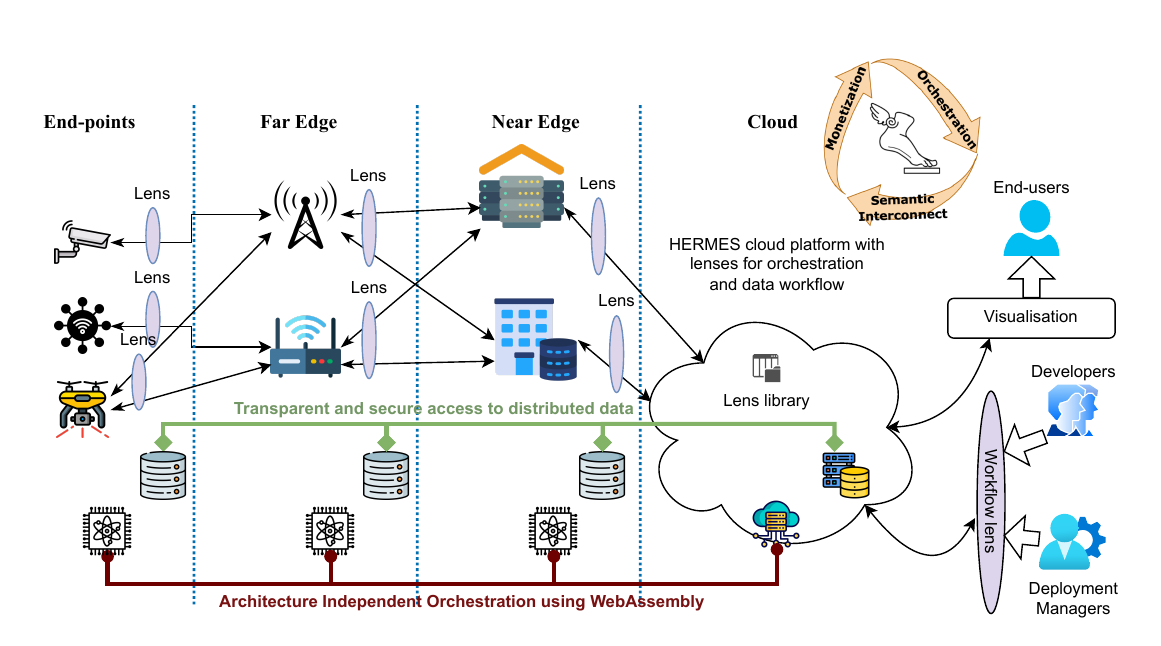}
\caption{Bird's-eye view of the HERMES architecture, showing cloud, edge, and IoT tiers unified by the three bridges: monetization, orchestration, and semantic interconnect.\label{fig:architecture}}
\end{figure*}

We present HERMES (Heterogeneous Computing Continuum with Resource Monetization, Orchestration, and Semantic) as a forward-looking vision to overcome these limitations and unlock the potential of a truly \emph{hyper-distributed} computing continuum. HERMES positions itself as a bridge between the current fragmented state of IoT and edge computing systems and a more cohesive, standardized future. By advancing technologies like WebAssembly (WASM)\footnote{WebAssembly, https://webassembly.org/} for infrastructureless orchestration, distributed ledgers, and named-data-access data fabric mechanisms for seamless and secure data access, as well as edge intelligence, HERMES aims to enable a cloud-edge continuum framework that allows applications to be deployed across multiple hardware architectures without modification. This approach not only facilitates easier
development and deployment of next-gen hyper-distributed applications but also combats vendor lock-in,
promoting a competitive and diverse market ecosystem. HERMES focuses on providing transparent and secure access to distributed data and AI service sharing, utilizing a novel concept called Lenses (as illustrated in Figure \ref{fig:architecture}), which leads to improved resource management and enables simpler monetization across various technology platforms. 
In the following, we describe the HERMES architecture and its core components, illustrate how it applies to real-world use cases, and discuss the challenges and opportunities on the horizon for this approach.

\section{HERMES Architecture}
At its core, HERMES defines an open architecture for the cloud-edge-IoT continuum that weaves together the three bridging concepts (as illustrated in Figure~\ref{fig:architecture}). These bridges address distinct but complementary aspects:
\begin{itemize}
\item \textbf{Monetization Bridge:} Establishes a decentralized marketplace for computing resources, data, and services across stakeholders.
\item \textbf{Orchestration Bridge:} Enables seamless deployment and management of applications on heterogeneous devices from cloud to edge.
\item \textbf{Semantic Interconnect:} Provides common data models and interfaces to ensure interoperability and meaningful information exchange.
\end{itemize}
HERMES connects these layers in a unified framework. We next detail each bridge and the key architectural components supporting it (as illustrated in Figure~\ref{fig:bridges}).

\subsection{Monetization: Distributed Marketplace and Incentives}
One of HERMES's novel aspects is treating \emph{everything-as-a-service} in the continuum, not only cloud computing power, but also edge device processing time, IoT sensor data feeds, and even trained AI models can be monetized. HERMES introduces a distributed marketplace, implemented with blockchain and smart contracts, where producers and consumers of such assets can trade in a secure, transparent manner~\cite{christidis2022decentralized}. For example, an industrial site with idle edge servers could offer spare computing cycles to nearby devices, or an autonomous vehicle fleet could sell real-time traffic sensor data to city authorities. The marketplace handles pricing, discovery, and trust between parties. To encourage participation, HERMES employs incentive mechanisms (e.g., tokens or credits) that reward nodes for contributing resources or sharing data. These economic incentives are designed to establish an equilibrium between the supply and demand of edge resources, transforming what were traditionally underutilized assets into active participants in the computing ecosystem.

Security and fairness are fundamental to the monetization bridge. By leveraging blockchain ledgers, transactions (e.g., a device purchasing computation from a nearby gateway) are recorded immutably, and cryptographic techniques ensure transactions and data exchanges remain tamper-proof~\cite{xue2023integration}. HERMES's marketplace smart contracts also embed policies, for instance, enforcing data usage agreements or ensuring payments are released only upon successful task completion. Overall, this bridge transforms the continuum into an \emph{economy of compute and data}, incentivizing organizations to cooperate and share resources in ways that were previously impossible.

\subsection{Orchestration: Seamless Deployment Across Tiers}
The orchestration bridge of HERMES tackles the complexity of managing applications across a diverse landscape of devices and networks. A key design principle is to be \emph{cloud-agnostic and edge-native} – rather than extending a monolithic cloud orchestrator to the edge. HERMES utilizes a hierarchy of lightweight orchestrators and intelligent agents that function across different tiers (i.e., cloud, fog, edge, and device) while coordinating with one another \cite{bartolomeo2023oakestra}. This allows global optimization goals (e.g., load balancing or energy efficiency) to be met while still respecting local constraints (e.g., real-time deadlines or battery limits) at the edge. To enable code to run anywhere in the continuum, HERMES makes extensive use of \textit{WASM} technology. WASM provides a portable binary format that can execute on any hardware platform via a lightweight runtime, essentially offering a \textit{"write once, run everywhere"} solution for heterogeneous environments \cite{menetrey2022webassembly}. In HERMES, application components can be compiled to WASM modules, shipped to edge nodes, and executed in sandboxed runtimes with minimal overhead \cite{gackstatter2022pushing}. This approach decouples services from the underlying hardware/OS – an important step toward an architecture-independent orchestration. In fact, even resource-constrained IoT devices can host WASM runtimes, allowing them to execute logic received from the cloud or other devices, much like containers but with far smaller footprints. 

\begin{figure}[h]
\centering
\includegraphics[width=\columnwidth]{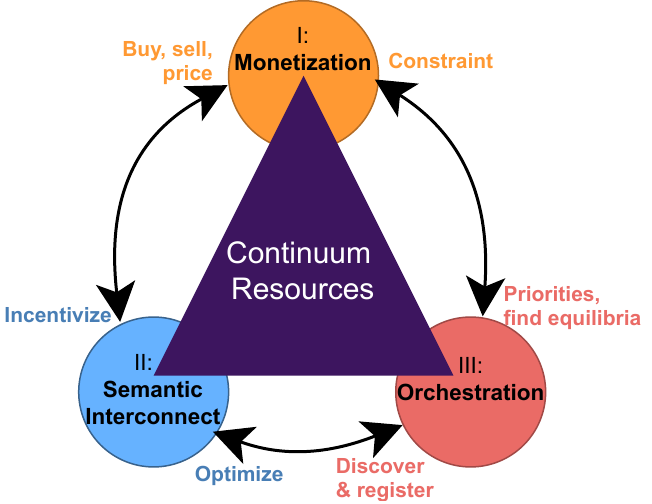}
\caption{The three primary HERMES bridges: monetization, semantic interconnect, and orchestration. Each bridge addresses a critical gap in today's edge-cloud systems, and together they enable a cohesive computing continuum.\label{fig:bridges}}
\end{figure}

Another key concept introduced by HERMES is that of \textit{lenses}. A lens is a pluggable data-processing module that can be dynamically inserted along the data path between IoT devices, edge nodes, and the cloud. By deploying lenses at strategic points, HERMES enables in-situ data filtering, aggregation, or analysis to occur closer to data sources when beneficial. For example, a video camera feed might have a "privacy lens" attached at an edge gateway to anonymize the video stream before it leaves a factory floor. Lenses are managed by the orchestration layer, meaning that they can be added, removed, or reconfigured on the fly in response to changing conditions or requirements. This gives HERMES flexibility, such as data can be pre-processed at the edge to reduce bandwidth, or enriched with local context, or split into multiple streams for different consumers. 

Orchestration also encompasses advanced scheduling and placement algorithms. Since not all edge nodes are equal (some may have GPUs, while others run on battery, etc.), the orchestrator uses a \emph{capability model} to match tasks to the appropriate nodes. The orchestrator takes into account factors like network latency, compute capacity, energy availability, and data locality. AI techniques (e.g., reinforcement learning) are planned to improve decision-making over time, learning the optimal distribution of workloads in the continuum \cite{ismail2025survey}. Crucially, orchestration decisions in HERMES are made with end-to-end goals in mind – e.g., minimizing overall response time of an IoT application or maximizing throughput – rather than being siloed at each tier. Lastly, HERMES incorporates \textit{security by design} in orchestration. A custom threat model, called DIRECTS (i.e., covering Denial-of-Service, Integrity, Repudiation, Elevation of Privilege, Confidentiality, Trust, and Spoofing), guides the design of all components. The orchestrator agents authenticate each other and the code they deploy; lens modules and WASM binaries are signed and verified; and data integrity is maintained via cryptographic hashes (potentially stored on a blockchain for audit). 

\subsection{Semantic Interconnect: Shared Knowledge and Interoperability}
The third bridge, semantic interconnect, addresses one of the thorniest issues in large-scale distributed systems: how to ensure that a multitude of devices and services \emph{understand} each other's data. In HERMES, all participants in the continuum – from tiny sensors to cloud analytics platforms – can publish and consume data through a common semantic description framework~\cite{ganzha2017semantic}. This builds upon concepts from the Semantic Web and recent IoT interoperability standards \cite{rahman2020comprehensive}. Essentially, HERMES defines ontologies and schemas for common IoT data types and contexts (e.g., sensor readings with time and location metadata, or actuator commands with units and expected effects). Edge devices annotate their data with these semantic descriptors, and edge/cloud services advertise their capabilities (e.g., "temperature forecasting service") in a machine-interpretable form.

HERMES enables dynamic discovery and integration of services by using semantic metadata \cite{gomes2019semantic, murturi2021decentralized}. For instance, if an application requires "air quality data within 5 km", the semantic broker in HERMES can locate any compliant data sources (i.e., regardless of who operates them or what protocol they use) and facilitate a connection (i.e., potentially via the monetization bridge if access requires payment or permission). This greatly reduces the manual effort typically needed to integrate across different IoT platforms. The semantic layer enhances the efficiency of data exchange, allowing consumers to subscribe to exactly the information they need based on high-level queries (e.g., SQL for IoT), and producers to filter or transform data into the required format via lenses. Moreover, the semantic interconnect also contributes to data governance and context awareness in HERMES. Because data is annotated with context (i.e., source device, quality, relationships to other data, etc.), policies can be enforced more intelligently. For instance, a rule might state that "video data from indoor cameras can only be processed by services that have a privacy certification". The orchestrator, informed by semantic tags on data and service descriptions, can then ensure compliance by only deploying certain tasks on authorized nodes.

In practical terms, HERMES implements a semantic registry and knowledge base accessible to all components. Edge nodes can cache relevant parts of the ontology for local decision-making (i.e., to avoid always contacting a cloud server). HERMES may utilize compressed knowledge graphs or distributed ledger entries to propagate key semantic information across the network efficiently. By creating a \emph{web of meaning} across the continuum, the semantic interconnect bridge allows heterogeneous components to work in concert, paving the way for more autonomous and intelligent behavior. Notably, this bridge also enables HERMES to incorporate higher-level concepts, such as \textit{digital twins}. Each physical entity (i.e., device, machine, or even a logical entity like a service) can have a semantic representation (i.e., a digital twin) that aggregates its state and interactions. Local "edge twins" can sync with more detailed cloud twins, providing real-time views and control of assets regardless of physical location. This synergy between digital twins and semantic interoperability further enhances the continuity of knowledge in the system. Lastly, Figure~\ref{fig:bridges} illustrates how the three bridges of HERMES relate to each other. Monetization and orchestration, together, enable market-driven resource management, while semantic interconnect underpins both by ensuring that what is being traded or orchestrated is well understood by all parties. 

\section{Use Cases and Validation}
To demonstrate the HERMES vision in action, consider several pilot scenarios drawn from different domains. These use cases demonstrate how the monetization, orchestration, and semantic interconnect bridges work together to enable capabilities that surpass those of current systems.

\subsection{Industry 4.0 – Predictive Maintenance}
\begin{figure}[h]
\centering
\includegraphics[width=\columnwidth]{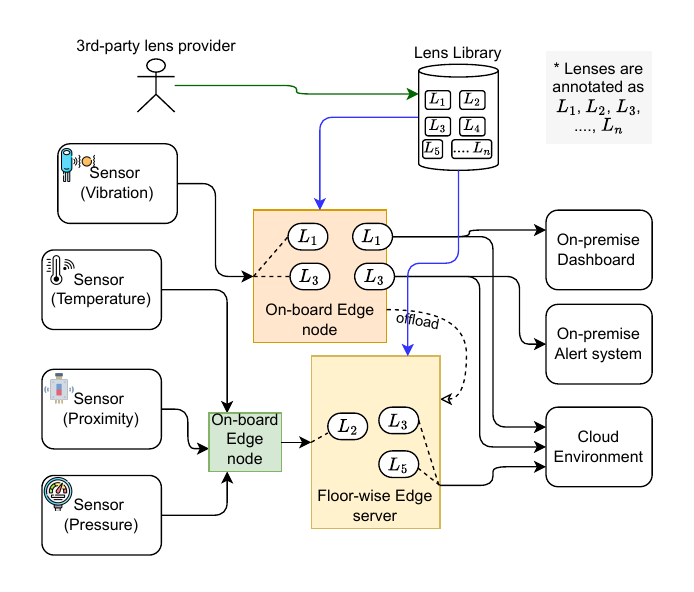}
\caption{Demonstration of HERMES, especially Lenses, in Industry 4.0 use case.}\label{fig:usecase1}
\end{figure}

Modern factories are increasingly equipped with IoT sensors on machines to monitor vibrations, temperatures, and other indicators of equipment health. However, in practice, much of this rich data is sent to cloud platforms for analysis long after critical events occur. Using HERMES, a manufacturing plant can push intelligence to the edge for real-time \emph{predictive maintenance}. Small edge servers on the factory floor (or even embedded compute modules on the machines) run HERMES lens modules that continuously analyze sensor streams for early signs of wear or anomalies. Lens modules can be taken from the lens library. Furthermore, it is not the responsibility of the corresponding industry/organization to develop the lenses; instead, these can be developed by any third party and shared through the library (as shown in Figure \ref{fig:usecase1}), encouraging their reuse. For instance, a vibration sensor on a turbine might have an AI inference lens that detects an emerging imbalance. If detected, the lens can immediately alert a local maintenance dashboard and also stream just the summarized anomaly information to the cloud (rather than raw high-bandwidth data).

The orchestration bridge ensures these analytics services are optimally placed. Suppose one of the machine's edge devices is overloaded. In that case, HERMES can dynamically offload some lens processing to a nearby device or a fog node at a higher level, while still maintaining low latency. Importantly, the semantic interconnect in HERMES ensures that all sensor data and alerts utilize a common vocabulary (e.g., vibration amplitude, temperature threshold), allowing higher-level systems (i.e., a cloud-based maintenance planning service) to easily aggregate and understand events from diverse machines. Through monetization mechanisms, equipment vendors or third-party service providers could also participate. For example, the factory's HERMES platform might allow an external specialist company to deploy a proprietary "failure prediction" lens on the factory's edge nodes for a fee, improving the accuracy of predictions using that company's algorithms. Thanks to HERMES's secure marketplace and unified architecture, this can be done without exposing raw data or compromising on latency. Figure~\ref{fig:usecase1} sketches this scenario: IoT sensors feed into on-site HERMES nodes, issues are detected at the edge, and maintenance actions are triggered proactively, reducing unplanned downtime.

\subsection{Autonomous Transportation – Smart Tram System}
Consider an urban deployment of autonomous trams or buses. These vehicles generate enormous amounts of data from sensors like cameras, LiDARs, and diagnostic systems. A key challenge is ensuring safety and operational efficiency in real-time, even when connectivity to the cloud is limited or intermittent. In a HERMES-enabled smart transportation pilot, each tram is equipped with an edge computing unit running HERMES. The orchestration bridge allows critical functions (e.g., object detection for obstacle avoidance) to execute on-board with minimal latency, while less urgent tasks (like aggregating ridership statistics or training an AI model on collected data) can be offloaded to wayside edge servers at stations or to the cloud when connectivity permits.


The semantic interconnect comes into play by standardizing the representation of different data sources. For example, a tram's HERMES node can semantically annotate events ("emergency brake applied," "signaling fault at location X") and share them in real-time with the city's traffic management center or even with other nearby vehicles. Because HERMES provides a common data model, a traffic management application can automatically integrate these feeds with data from other sources (e.g., smart traffic lights or sensors on the tracks) to get a holistic view of the situation.

 Trust and safety are paramount in this scenario. HERMES leverages its security features (i.e., blockchain-backed data integrity and the DIRECTS threat model) to ensure that data exchanged between trams, roadside infrastructure, and cloud services cannot be tampered with. For instance, maintenance logs or safety-critical messages can be notarized on a distributed ledger, allowing authorities to verify their authenticity and integrity. Moreover, the monetization bridge may allow new stakeholder interactions: imagine a scenario where the tram operator's HERMES system offers anonymized telemetry data to third parties (such as researchers or urban planners) under a paid data-sharing agreement, creating an additional revenue stream. HERMES handles the enforcement of such agreements by only releasing the data (via semantic queries) to the buyer's services once the smart contract conditions are met. Overall, in the autonomous transportation use case, HERMES provides resilience (e.g., local autonomy when the cloud is unavailable), situational awareness (through semantic data sharing), and a framework for secure collaboration between public and private entities.

\subsection{Precision Agriculture: Smart Farming}

Farmers increasingly deploy IoT sensors, drones, and AI models to boost yields while conserving resources. HERMES enhances such smart farming ecosystems through seamless orchestration of devices and data. A rural farm might utilize soil moisture sensors, weather stations, and camera drones, all of which are managed by an edge gateway. When a drone detects pests or crop diseases, an image-analysis lens (i.e., possibly from an agri-tech provider) runs locally or on the farm's edge server. Detected issues are labeled with geotagged semantic annotations (e.g., "fungus in north field, two hectares affected"). The semantic interconnect ensures all annotations use standard vocabularies understood by farm management tools and advisory systems.

Reliable operation is critical in rural regions with intermittent connectivity. The local HERMES network stores data, triggers autonomous actions such as targeted irrigation, and synchronizes with cloud analytics when links are restored. Summarized data (e.g., soil health, growth metrics) are transmitted to the cloud for comparison against regional trends. The orchestration bridge reallocates workloads as needed. If a drone's battery level drops, processing is shifted to the gateway or deferred until the drone is uploaded.

Monetization emerges through cooperative data exchange among farmers, researchers, and service providers. Farms can form a data-sharing consortium, where each contributes anonymized data to a collective pool that improves regional pest models, and in return, gains access to enhanced analytics. Smart contracts on the HERMES marketplace automate exchanges and ensure fairness; farms providing more data earn greater rewards. The same framework supports buying auxiliary data, such as high-resolution weather forecasts, directly through the marketplace. Unified data formats and automation lower technical barriers, allowing even small farms to access advanced digital agriculture services. HERMES enables smart farms to connect and share intelligence, promoting resilient and sustainable agriculture.

\section{Future Outlook and Challenges}
HERMES approach is ambitious in scope, and its realization will require overcoming several open challenges. One major area is the development of {open standards} for the computing continuum. For HERMES to achieve widespread adoption, its protocols for resource trading, semantic annotation, and orchestration must be standardized, allowing different vendors' devices and platforms to interoperate. 

\subsubsection{Scalability and performance} Another challenge lies in {scalability and performance}. Operating a distributed marketplace and orchestrator across potentially millions of devices (some of which are very constrained in terms of power and bandwidth) is non-trivial. Techniques such as hierarchical federation of marketplaces and multi-level orchestrators will be necessary to enable decisions to be made locally when possible and globally when necessary, without overwhelming the network or control systems. The use of AI in orchestration (e.g., learning optimal deployment strategies) also raises questions about how to provide guarantees. Future research must ensure that such AI-driven management remains predictable and avoids instability in highly dynamic environments.

\subsubsection{Security and trust} HERMES integrates blockchain, advanced cryptography, and the DIRECTS threat model to enhance security. The evolving threat landscape (including future quantum computer capabilities) means continuous updates will be required. One intriguing direction is to explore {self-sovereign identity} and trust frameworks for devices, so that each IoT node in HERMES can carry its own verifiable credentials and policies wherever it roams in the continuum. Ensuring privacy, especially in the context of a data marketplace, is another challenge. Techniques such as secure multi-party computation or federated learning may be leveraged in future HERMES iterations to enable collaborative analytics without exposing raw data.

\subsubsection{Open challenges} On the research front, HERMES opens many questions. How can economic incentive models be balanced with technical efficiency? For example, could a surge in cryptocurrency prices inadvertently incentivize too many edge devices to perform mining instead of their primary IoT functions? How will semantic interoperability scale as ontologies grow and diversify across domains? There is a need for robust {governance models} for the continuum: essentially, frameworks to manage policy (who can deploy what lens where, how disputes in marketplace transactions are resolved, etc.) in a decentralized way. This may involve a combination of on-chain governance for marketplace rules and off-chain agreements for regulatory compliance.

Despite these challenges, the trends in technology strongly align with the HERMES vision. Computing is increasingly moving towards the edge, AI/ML are becoming ubiquitous in distributed forms, and there is a growing recognition that data has value that can be transacted. The next few years may see the early implementation of HERMES principles. For instance, experimental deployments in smart cities or industrial testbeds that validate the performance of HERMES's marketplace or the effectiveness of lenses in reducing latency and bandwidth. Each success will pave the way for broader adoption. In the long term, if frameworks like HERMES mature, we could see a paradigm shift: rather than today's cloud-centric IoT, a true \emph{computing continuum} would emerge where intelligence, storage, and value exchange flow freely from tiny sensors up to mega-scale cloud centers.

\section{Conclusion}
HERMES represents a bold vision for the future of distributed computing. By unifying monetization, orchestration, and semantics, it aims to transform a fragmented edge and IoT landscape into a cohesive, collaborative continuum. In this article, we outlined how HERMES's framework can enable new capabilities (i.e., from marketplaces for edge resources to plug-and-play interoperability of services) that address longstanding pain points in IoT and cloud computing. Much work remains to implement and standardize these ideas, but the potential benefits are immense, including more efficient resource utilization, accelerated innovation through shared data and services, and greater autonomy for systems that can operate intelligently from the cloud to the edge.

\def\refname{REFERENCES}

 \bibliographystyle{ieeetr}
\bibliography{ref} 

\begin{IEEEbiography}{Chinmaya Kumar Dehury}{\,}is currently an Assistant Professor in Computer Science department, IISER Berhampur, India. He was an Assistant Professor in the Institute of Computer Science, University of Tartu, Estonia. His research interests include scheduling, resource management and fault tolerance problems of Cloud and fog Computing, the application of artificial intelligence in cloud management, edge intelligence, Internet of Things, and data management frameworks. His research results are published by top-tier journals and transactions such as IEEE TCC, JSAC, TPDS, FGCS, etc. He is a member of IEEE and ACM India. He is also serving as a PC member of several conferences and reviewer to several journals and conferences, such as IEEE TPDS, IEEE JSAC, IEEE TCC, IEEE TNNLS, Wiley Software: Practice and Experience, etc \vadjust{\vfill\pagebreak}
\end{IEEEbiography}

\begin{IEEEbiography}{Lauri Lovén} {\,} (SM'19), Assistant Professor (tenure track), is a PI and the vice director of the Center for Ubiquitous Computing (UBICOMP), University of Oulu, in Finland. He leads the Future Computing Group of ca. 20 researchers and coordinates the Distributed Intelligence strategic research area in the national 6G Flagship research program. His team studies future generation computing in the IoT-edge-cloud continuum, focusing on the theory and applications of agentic and edge intelligence.
\end{IEEEbiography}

\begin{IEEEbiography}{Praveen Kumar Donta}{\,} currently Associate Professor (Docent) at the Department of Computer and Systems Sciences, Stockholm University, Sweden. He received his Ph.D. from the Indian Institute of Technology (Indian School of Mines), Dhanbad, in the Department of Computer Science \& Engineering in June 2021. He is a Senior Member of the IEEE Computer and Communications Societies, and a Professional Member of the ACM. His research includes learning in Distributed Continuum Systems. Contact him at \texttt{praveen@dsv.su.se}. \vspace*{8pt}
\end{IEEEbiography}

\begin{IEEEbiography}{Ilir Murturi}{\,} is currently an Assistant Professor in the Department of Mechatronics, University of Prishtina, Kosova.   His research interests include the Internet of Things, Distributed Computing
Continuum Systems, and EdgeAI. Murturi received his Ph.D.
degree in Information Systems Engineering from TU Wien. He
is a Senior Member of IEEE. Contact him as the corresponding author at \texttt{ilir.murturi@uni-pr.edu}.\vspace*{8pt}
\end{IEEEbiography}

\begin{IEEEbiography}{Schahram Dustdar}{\,}  is a full professor of computer science and heads the Research Division of Distributed Systems at TU
Wien, 1040, Vienna, Austria, and UPF ICREA, 08018, Barcelona,
Spain. His research interests include the investigation of all aspects related to edge computing, fog computing, and cloud computing. Dustdar received his Ph.D. degree in Business Informatics from Johannes Kepler University Linz. He is a Fellow of IEEE. Contact him at dustdar@dsg.tuwien.ac.at. \vspace*{8pt}
\end{IEEEbiography}
\end{document}